\documentclass{article}
\usepackage[T1]{fontenc}
\usepackage[utf8]{inputenc}
\usepackage{spconf,amsmath,epsfig}

\let\OLDthebibliography\thebibliography
\renewcommand\thebibliography[1]{
  \OLDthebibliography{#1}
  \setlength{\parskip}{0pt}
  \setlength{\itemsep}{0pt plus 0.3ex}
}

\usepackage[colorlinks,
            linkcolor=blue,
            anchorcolor=blue,
            citecolor=blue]{hyperref}

\usepackage{booktabs}
\begin{document}\sloppy

% Example definitions.
% --------------------
\def\x{{\mathbf x}}
\def\L{{\cal L}}

% Title.
% ------
\title{GRNN:Recurrent Neural Network based on Ghost Features for Video Super-Resolution}
%
% Single address.
% ---------------
\name{Yutong Guo}
%Address and e-mail should NOT be added in the submission paper. They should be present only in the camera ready paper. 
\address{{Department of Electronics and Communication Engineering,Shanghai, China} \\
{East China University of Science and
Technology,Shanghai, China} \\
19001822@mail.ecust.edu.cn}

\maketitle

\begin{abstract}
Modern video super-resolution (VSR) systems based on convolutional neural networks (CNNs) require huge computational costs. The problem of feature redundancy is present in most models in many domains, but is rarely discussed in VSR. We experimentally observe that many features in VSR models are also similar to each other, so we propose to use "Ghost features" to reduce this redundancy. We also analyze the so-called "gradient disappearance" phenomenon generated by the conventional recurrent convolutional network (RNN) model, and combine the Ghost module with RNN to complete the modeling on time series. The current frame is used as input to the model together with the next frame, the output of the previous frame and the hidden state. Extensive experiments on several benchmark models and datasets show that the PSNR and SSIM of our proposed modality are improved to some extent. Some texture details in the video are also better preserved.
\end{abstract}
\section{Introduction}
In recent years, with the video platform, the rapid development of the field of live, people's attention to the video super-resolution(VSR) is also increasing. Unlike the  single image super-resolution(SISR) , VSR can use inter-frame information to improve the performance of the model. According to existing research,  the use of inter-frame information is divided into two ways: one is based on alignment, and the common methods include using optical flow\cite{bib32} to do motion estimation compensation, and using deformable convolution\cite{bib33} to do alignment in two categories. One is based on non-alignment , and four types\cite{bib38} of models are commonly used: 2DConv\cite{bib34}, 3DConv\cite{bib35}, Recurrent CNN\cite{bib36}, and Non-Local Network \cite{bib37}.

The Motion estimation compensation inside the first type of alignment method is done using optical flow. The purpose of motion estimation is to extract inter-frame motion information, while motion compensation performs inter-frame warping operations to align the inter-frame motion information. The common point of this class of hyper-segmentation algorithms is to align adjacent images with the target image using motion estimation and motion compensation techniques, but neither of them can guarantee the accuracy of motion information, especially when there are light changes or large movements, artifacts can easily appear\cite{bib39}. Deformable alignment is similar to optical flow alignment in that it also employs spatial deformation. In fact, the difference between deformable alignment and optical flow is the offsets (offsets). Instability can occur in the training of variable convolution parameters, resulting in the inability to learn valid offsets, which in turn affects the final result. The diversity of offsets is the main reason for the improved results for deformable alignment, but in practice we need to choose an appropriate amount, otherwise it will cause unnecessary computations for the network\cite{bib39}.

\begin{figure}[htp]
    \centering
    \includegraphics[width=8.5cm]{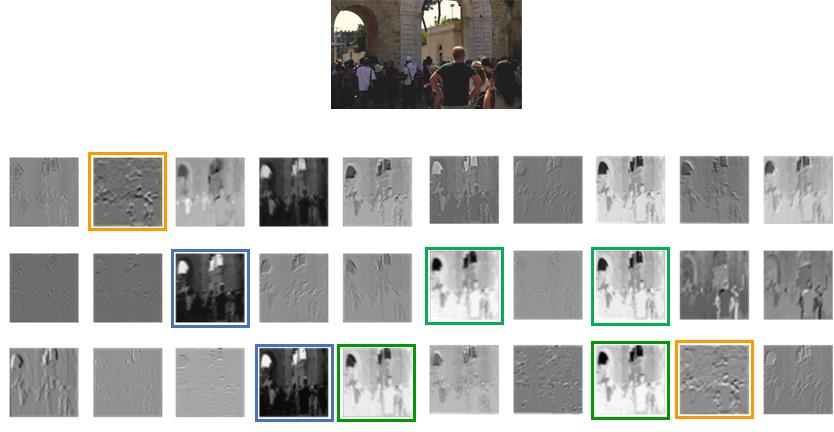}
    \caption{Similar feature maps where the rectangular boxes of the same color are similar to each other}
    \label{fig:galaxy}
\end{figure}

In contrast to the aligned methods, the non-aligned methods do not align adjacent frames of the video super-resolution. This type of method mainly uses spatial or spatio-temporal information for feature extraction. With the help of advanced temporal modeling strategies, CNN-based methods show excellent performance on several benchmarks. However, the overlapping of sliding windows leads to redundant computations, which limits the VSR efficiency\cite{bib20}. From our preliminary experimental results(Fig.1), there is a strong similarity in the feature maps generated after convolution. So we hope we can simplify these feature maps with some simple According to the experimental results of GhostNet\cite{bib18}, for the above-mentioned case of feature map redundancy, more similar feature maps can be generated by performing a simple linear operation on one of them, so that more feature maps can be generated with fewer parameters, and the similar feature maps can be considered as Ghosts of each other.

According to the existing studies (published ones) do not apply the above mentioned ideas of Ghost to the treatment of time series. Most of the studies on Ghost have been conducted only in a few fields. And, most of the studies so far have been descriptive in nature. In this paper, we combine this "Ghost" idea with RNN to simplify the feature maps of time series. Experimental results demonstrate the effectiveness of this approach, with improvements over existing PNSR and SSIM methods on all three experimental sets (Vid4\cite{bib22}, SPMCS\cite{bib24} and UDM10\cite{bib23}).

The main contributions of this paper are threefold.
(1) We propose a new time series that introduces Ghost's idea into convolutional recurrent networks, thus reducing the redundancy of the network.
(2) We design a residual module that skips the residuals between connected layers. Such a design ensures smooth information flow and has the ability to retain texture information for a long time, thus making GRNN easy to handle longer sequences while reducing the risk of gradient disappearance during training.
(3) Extensive experiments are conducted on three public benchmark datasets (i.e., Vid4, SPMCS, and UDM10), and the results show that our GRNN outperforms the state-of-the-art methods.

The following paper will introduce our work in terms of related works, approach, experiment and conclusion.

\section{Related Works}\label{sec2}
\subsection{Video super-resolution algorithms}\label{subsec2}
Deep learning based video super-resolution algorithms generally use convolutional neural networks (CNN)\cite{bib4}, generative adversarial networks (GAN)\cite{bib5}, or recurrent neural networks (RNN)\cite{bib6}. The architectures are basically using low resolution as input, then inter-frame alignment, feature extraction, feature fusion, and finally reconstruction to generate high resolution video. The biggest difference between video super-resolution and image super-resolution is that video super-resolution uses inter-frame information. How to use this information
efficiently is also where the different algorithms differ.
A simple classification of super-resolution algorithms is based on the method of using the information between adjacent frames: two categories of adjacent frames for alignment and non-alignment. 

\subsubsection{Adjacent frames for alignment}\label{subsec2}
Alignment algorithms can be further divided into two categories using motion estimation and motion compensation (MEMC) and using variable convolution.

\textbf{Using MEMC:}Motion estimation and compensation algorithms have a very important role in video super-resolution, and many algorithms are based on them. Motion estimation is to extract the motion information between frames and then align the different frames according to the motion information. Most of the motion estimation uses optical flow method, i.e., the motion information is obtained by calculating the time-domain correlation and variation between frames, and the commonly used methods are linear interpolation and space transform network (STN)\cite{bib7}.The main video super-resolution algorithms using MEMC are VSRnet\cite{bib16},VESPCN\cite{bib17}.

\textbf{Using variable convolution:}
Variable convolution was proposed in 2017 and differs from traditional convolution layers at the point that traditional convolution layers, each layer has a fixed size kernel; variable convolution adds an offset to the kernel so that the input features can be better transformed to the geometric model by convolution operations. The main video super-resolution algorithms using variable convolution are EDVR\cite{bib8}, DNLN\cite{bib9}, TDAN\cite{bib10}, D3Dnet\cite{bib11}, and VESR-Net\cite{bib12}.

\subsubsection{Adjacent frames for non-alignment}\label{subsec2}
In addition to the aforementioned alignment methods, there are various non-alignment algorithms, i.e., reconstruction without alignment operations on the frames. They can be subdivided into two-dimensional convolution method (FFCVSR)\cite{bib13}, three-dimensional convolution method (dynamic upsampling filtering (DUF)\cite{bib14}, cyclic convolution method (bidirectional recurrent convolutional network (BRCN)\cite{bib15}, and non-local network method. Except for the two-dimensional convolution method, all the methods use joint information in the space-time domain. These methods rely on neural network learning to obtain feature and motion information, and thus do not require frame alignment

\subsection{GhostNet: More Features from Cheap Operations}\label{subsec2}
In a well-trained deep neural network, it usually contains rich or even redundant feature maps to ensure a comprehensive understanding of the input data. GhostNet introduces a new Ghost block that aims to generate more feature maps through inexpensive operations. Based on a set of original feature maps, a series of linear transformations are used to generate many "ghost feature" maps that uncover the required information from the original features at a small cost.

\begin{figure}[htp]
    \centering
    \includegraphics[width=9cm]{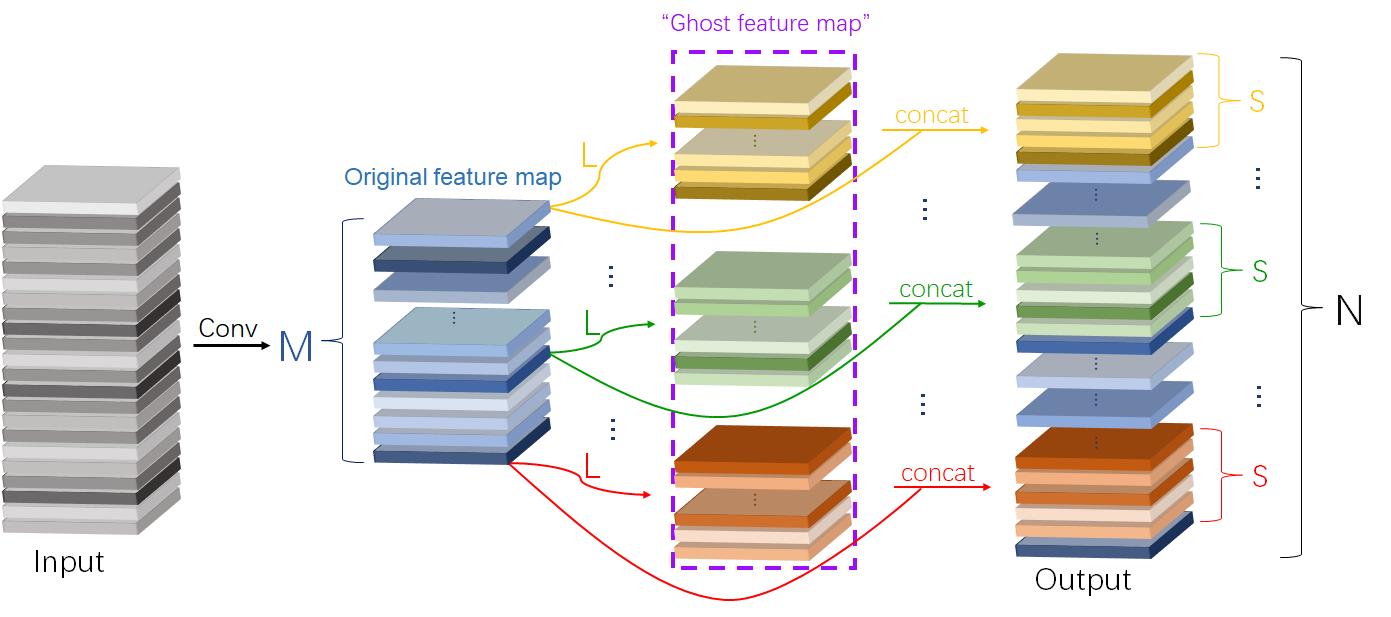}
    \caption{Flowchart of generating Ghost Block}
    \label{fig:galaxy}
\end{figure}
As shown in Fig. 2, the image is input to the Ghost block and then M original feature maps are generated by a normal convolution operation, and each original feature map is generated by a cheap linear operation L to generate the Ghost feature map(in the purple dashed box).The ghost feature map and the original feature map are output from the Ghost block after a concat operation to generate N feature maps.The ghost feature map and the original feature map are then concatenated to generate N feature maps. Then they output from the Ghost block.

Denote the original feature map as y and the Ghost feature map as y', then we get:
\begin{equation}
y_{i j}{\prime}=L_{i, j}(y_i), \quad \forall i=1, \ldots, M, \quad j=1, \ldots, S  \qquad \label{eq1}    
\end{equation}

$y_{i}$, denotes the ith original feature map and $y_{i j}{\prime}$,  denotes the jth ghost feature map of $y_i$, where $j$ can be equal to 1. Generate $N = M\times S$ feature maps $Y=[y_{1 1}{\prime},y_{1 2}{\prime},...,y_{i j}{\prime}]$. Since the linear operation $L$ is computed on each channel, it is much less computationally intensive than the ordinary convolution.

\section{Approach}\label{sec3}
By observing the feature maps output by the RNN model for a given video, we find that the feature maps inside also have some similarity with each other,as shown in Fig. 1.
Based on the above phenomenon, we combine the idea of Ghost with Recurrent Neural Network (RNN) to realize the modeling of a given video sequence in time.
The classical RNN model is a class of neural networks with short-term memory capability. In a recurrent neural network, neurons can receive information not only from other neurons, but also from themselves, forming a loop structure. In many real-world tasks, the output of the network is related not only to the current input but also to the output of a past period.
\begin{figure}[htp]
    \centering
    \includegraphics[width=9cm]{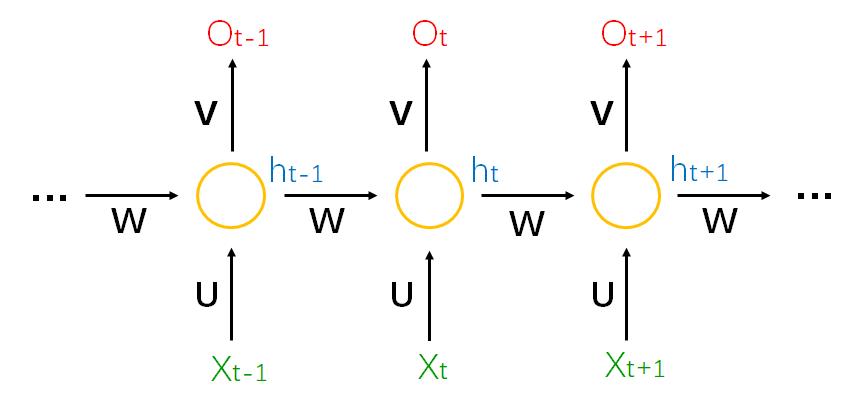}
    \caption{Flowchart of RNN}
    \label{fig:galaxy}
\end{figure}

As shown in Fig. 3,After this network receives input $X_t$ ,  at time t, the value of the hidden layer is $h_t$, and the output value is $O_t$. The key point is that the value of $h_t$ depends not only on $X_t$, but also on $h_{t-1}$. The above relationship is expressed in mathematical form as follows: 
\begin{equation}
O_t=g\left(V \cdot h_t\right) \label{eq1}
\end{equation}
\begin{equation}
h_t=f\left(U \cdot X_t+W \cdot h_{t-1}\right)\label{eq1}
\end{equation}
Most convolutional networks can only take and process one input individually, and the previous input is completely unrelated to the latter one. But when we process video, we need to be able to process the information of the sequence better, so we can't just analyze each frame individually, but the whole sequence of these frames connected together.Therefore, RNN, which is a very effective model for processing data with sequence characteristics, is very suitable for processing video sequences.Fig. 4 shows the relevant code module of RNN in processing video sequences.For each time t, the inputs are: (i) the output $O_{t-1}$, at moment $t-1$, (ii) the hidden state $h_{t-1}$, at moment $t-1$, (iii) the current frame $F_t$, and the previous frame $F_{t-1}$.

\begin{figure}[htp]
    \centering
    \includegraphics[width=9cm]{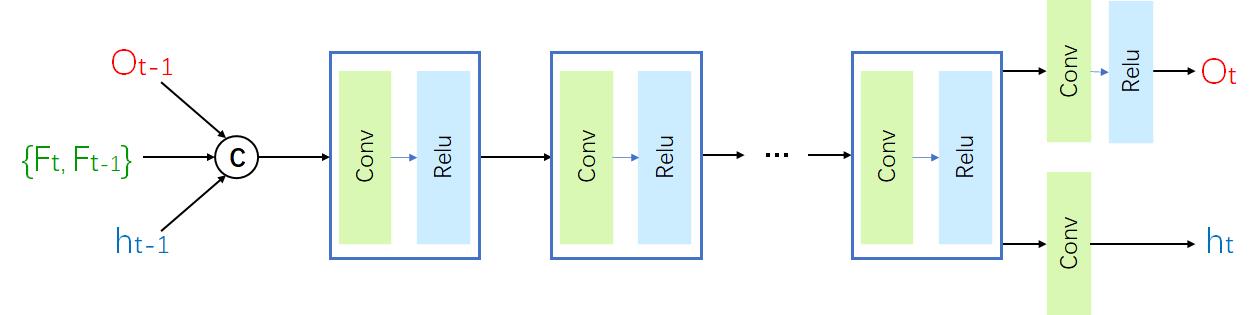}
    \caption{Relevant code module of RNN}
    \label{fig:galaxy}
\end{figure}
Since the weights of RNN are shared, as shown in Fig. 3, the V gradient at a certain moment will not be a problem, but W and U, because each moment is determined by all the previous moments together, so when the distance is long,the leading derivative will disappear or explode, but the overall gradient of the current moment will not disappear, because it is a summation process. The gradient of RNN does not disappear, “ the gradient of RNN disappears” means that the current gradient is not used by the previous gradient,  the gradient is dominated by the near gradient, which makes it difficult for the model to learn the dependencies at a long distance.

In our work, we retain the form of input and output shown in Fig. 4, and add a Ghost module to the original RNN to reduce the computational effort of the model. We also improve the RNN model structure by using a single residual block\cite{bib19}\cite{bib20} with skip connections to solve the problem that it is difficult to learn long-range dependencies in the previous RNN model.We named our model GRNN.
\begin{figure*}[htp]
    \centering
    \includegraphics[width=17.6cm]{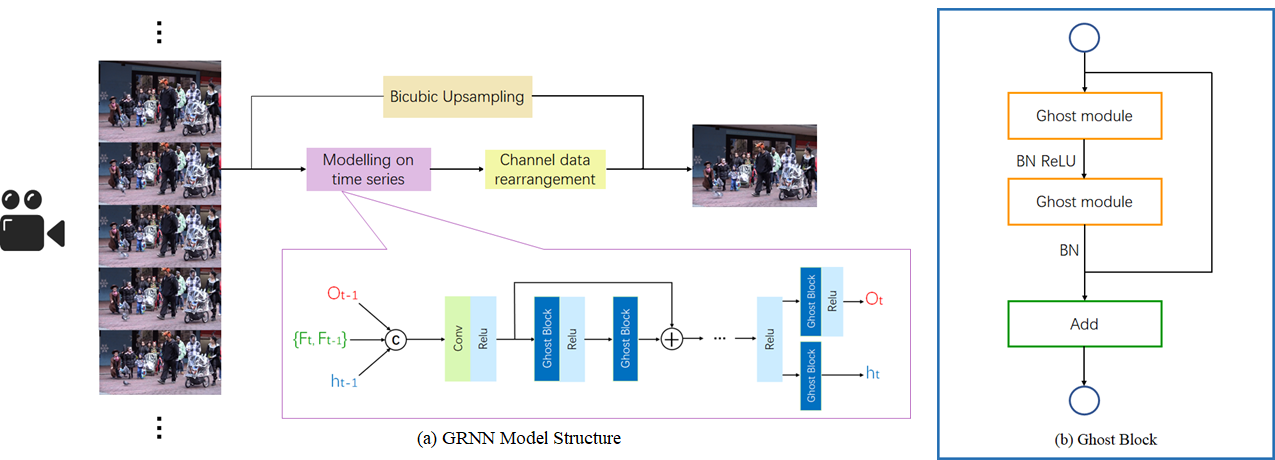}
    \caption{ GRNN Model Structure. The  box (a) shows the overall design idea of the model, and the internal structure of Ghost Block is the blue box (b).
}
    \label{fig:galaxy}
\end{figure*}

\section{Experiment}\label{sec4}
\subsection{Experiment Dataset}\label{subsec2}
 In the training phase of the model, the dataset we used is the Vimeo-90k dataset, a large-scale, high-quality video dataset for processing low-level video, provided at the same time as the TOFlow model presented in \cite{bib21} thesis. The dataset contains 89,800 video clips downloaded from vimeo.com, covering a variety of scenes and actions, and can be used to solve four video processing tasks: video interpolation, video noise reduction, video unlocking, and video super-resolution.And the dataset is processed as follows: $64\times64$ LR frames are generated from $256\times256$ by Gaussian blur with standard deviation $\sigma = 1.6$ and 4x downsampling.

 During the testing phase, we used Vid4\cite{bib22}, SPMCS\cite{bib24}, and UDM10\cite{bib23}, which are commonly used datasets in the field of video super-resolution.
 
\subsection{Experiment  Details}\label{subsec2}
We employ 10  residual blocks as GRNN for the implied states. each block consists of a convolutional layer, a ReLU layer, and another convolutional layer. The channel size of the convolutional layer was set to 128. At time step $t_0$, the previous estimates were initialized to zero. The learning rate was initially set to $1 \times 10^{-4}$, then reduced by a factor of 0.1 every 10 epochs until 70 epochs. All models were supervised by pixel-level $L_1$ loss functions and Adam \cite{bib25} optimizer with settings
$\beta_1 = 0.9$, $\beta_2 = 0.999$ and weight decay of $5 \times 10^{-4}$. We set the small batch sizes to 64 and 4, respectively. the $L_1$ loss was applied to all pixels.
All experiments were performed using Python 3.6.4 and Pytorch 1.1.
\subsection{Experiment  Results}\label{subsec2}

In this section we will show some experimental results as detailed in Table 1, Table 2, Fig. 6 and Fig. 7.

\begin{figure}[htp]
    \centering
    \includegraphics[width=8.3cm]{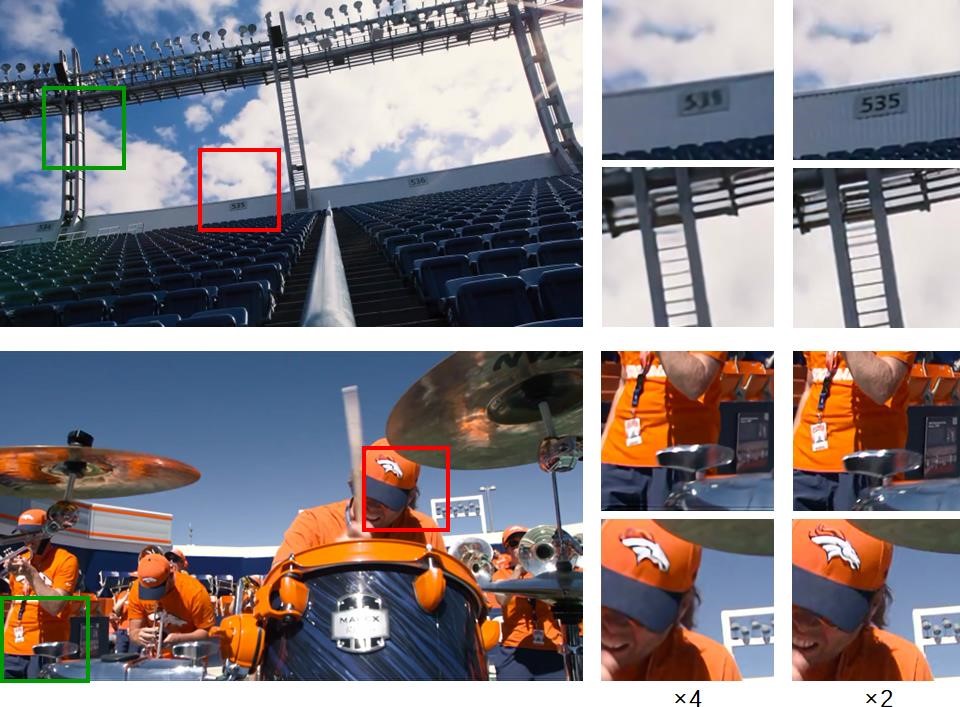}
    \caption{ Qualitative results of upscaling the 4× VSR and 2×VSR downscaled images.  
}
    \label{fig:galaxy}
\end{figure}

\begin{table*}[h]
\vspace{-2.0em}
\begin{center}

\begin{minipage}{\textwidth}
%%\caption{Example of a lengthy table which is set to full textwidth}\label{tab2}
\caption{Quantitative comparison(PSNR(dB)/SSIM) on Vid4\cite{bib22} for 4×VSR.‘†’ means the values are taken from original publications or calculated by provided models.Red text indicates thebest and blue text indicates the second best performance.}\label{tab2}
\begin{tabular*}{18cm}{@{\extracolsep{\fill}}lccccc@{\extracolsep{\fill}}}

\toprule%

Method(RGB) &  Calender & City & Foliage & Walk & Average\\
\midrule
Bicubic  & 17.04/0.4616 & 22.05/0.4914 & 19.74/0.4118 & 22.65/0.6776 & 20.37/0.5106\\
SPMC†\cite{bib26}  & -/- & -/- & -/- & -/- & 25.52/0.7600\\
Liu†\cite{bib27}  & 21.61/ -  & 26.29/ -  & 24.99/ -  & 28.06/ -  & 25.23/ - \\
TOFlow†\cite{bib21}  & 20.83/0.7062 & 25.34/0.7255 & 23.84/0.6967 & 27.55/0.8468 & 24.39/0.7438\\
FRVSR†\cite{bib29}  & - & - & - & - & 25.01/0.7917\\
DUF† \cite{bib14}  & \textcolor{blue}{22.67}/0.8062 & \textcolor{blue}{26.89}/0.8066 & 24.96/\textcolor{blue}{0.7681} & \textcolor{blue}{29.12}/0.8855 & \textcolor{blue}{25.91}/0.8166\\
RBPN† \cite{bib30}  & 22.32/0.7597 & 26.25/0.7801 & \textcolor{blue}{25.35}/0.7581 & 28.67/0.9009 & 25.65/0.7997\\
EDVR-L† \cite{bib8}  & 22.64/0.8046 & 26.76/0.7986 & 24.84/0.7442 & 29.08/0.8834 & 25.83/0.8077\\
PFNL† \cite{bib31}  & 22.43/\textcolor{blue}{0.8101} & 26.62/\textcolor{blue}{0.8107} & 24.87/0.7513 & 28.76/\textcolor{red}{0.9036} & 25.67/\textcolor{blue}{0.8189}\\
GRNN(ours)  & \textcolor{red}{23.59}/\textcolor{red}{0.8102} & \textcolor{red}{27.77}/\textcolor{red}{0.8161} & \textcolor{red}{26.02}/\textcolor{red}{0.7696} & \textcolor{red}{29.94}/\textcolor{blue}{0.9031} & \textcolor{red}{26.83}/\textcolor{red}{0.8248}\\
\bottomrule
\end{tabular*}
%\footnotetext[1]{Example for a first table footnote.}
%\footnotetext[2]{Example for a second table footnote.}
\end{minipage}
\end{center}
\end{table*}

\begin{figure*}[htp]
\vspace{-1.0em}
    \centering
    \includegraphics[width=17.8cm]{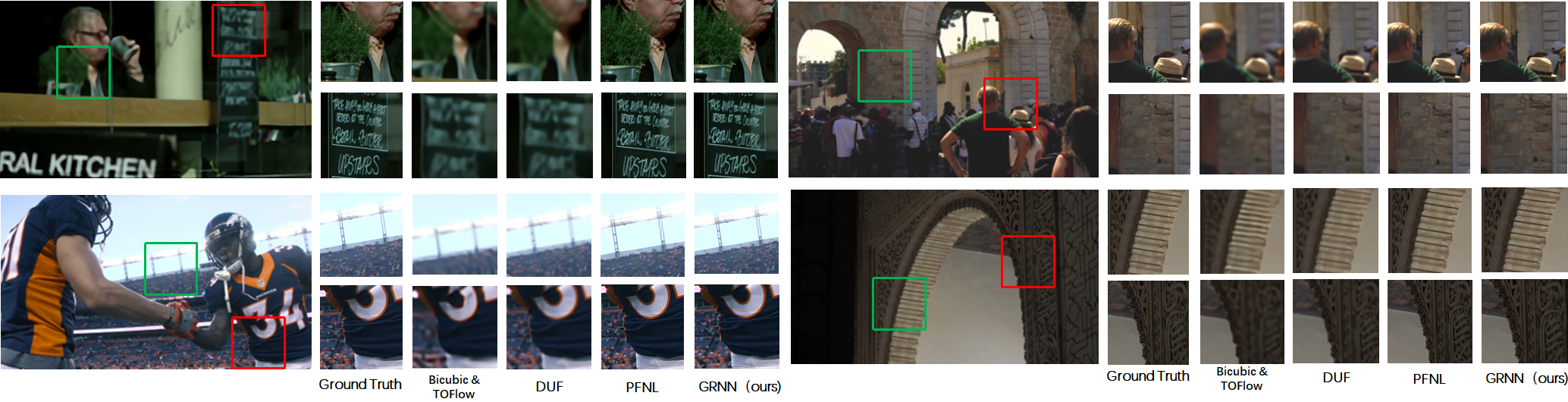}
    \caption{ Qualitative results of upscaling the 4× VSR downscaled images. GRNN recovers rich details,
leading to both visually pleasing performance and high similarity to the original images. 
}
    \label{fig:galaxy}
\end{figure*}

\begin{table}[htp]
\vspace{-1.0em}
\caption{Quantitative comparison(PSNR(dB)/SSIM) 
on Vid4\cite{bib22}, SPMCS\cite{bib24}, and UDM10\cite{bib23} for 4×VSR,and the performance of our model with 2×VSR}\label{tab2}
\vspace{-2.0em}
\begin{center}
\begin{minipage}{\textwidth}
\begin{tabular*}{9cm}
{@{\extracolsep{\fill}}lccc@{\extracolsep{\fill}}}
\toprule
Method & Vid4\cite{bib22} & SPMCS\cite{bib24} & UDM10\cite{bib23} \\
\midrule
Bicubic   & 20.37/0.5106 & 21.83/0.6133 & 27.05/0.8267\\
TOFlow†\cite{bib21}   & 24.39/0.7438 & 26.38/0.8072 & 34.46/0.9298\\
FRVSR†\cite{bib29}   & 25.01/0.7917 & 26.68/0.8271 & 35.39/0.9403\\
DUF† \cite{bib14}   & \textcolor{blue}{25.91}/0.8166 & 28.10/\textcolor{blue}{0.8582} & \textcolor{blue}{36.78}/0.9514\\
RBPN† \cite{bib30}   & 25.65/0.7997 & \textcolor{blue}{28.23}/0.8561 & 36.53/0.9462\\
EDVR-L† \cite{bib8} & 25.83/0.8077 & -/- & -/-\\
PFNL† \cite{bib31}   & 25.67/\textcolor{blue}{0.8189}   & 27.24/0.8495 & \textcolor{red}{36.91}/\textcolor{blue}{0.9526}\\
GRNN(×4)  & \textcolor{red}{26.83}/\textcolor{red}{0.8248} & \textcolor{red}{28.34}/\textcolor{red}{0.8631} & 36.35/\textcolor{red}{0.9548} \\
GRNN(×2) & 33.42/0.9535 & 35.52/0.9705 & 44.28/0.9922\\
\bottomrule
\end{tabular*}
%\footnotetext[1]{Example for a first table footnote.}
%\footnotetext[2]{Example for a second table footnote.}
\end{minipage}
\end{center}
\vspace{-3.0em}
\end{table}

Based on the above qualitative analysis, our model GRNN, demonstrates superior performance on most of the datasets.

%%=============================================%%
%% For presentation purpose, we have included  %%
%% \bigskip command. please ignore this.       %%
%%=============================================%%

%%=============================================%%
%% For presentation purpose, we have included  %%
%% \bigskip command. please ignore this.       %%
%%=============================================%%

\section{Conclusion}\label{sec13}
We found the similarity between feature maps by observing the feature maps generated by the RNN-based network, which can be obtained by cheap linear convolution, so we introduced Ghost Block. in the RNN model to reduce the redundant feature maps generated by the model. We also added a residual module in the second half of the model to alleviate the phenomenon of "gradient disappearance" that often occurs in RNN models. The final result is to model the Ghost idea on  time series. Our model shows good performance on most of the datasets. The video quality is also clearer in terms of visual intuition.
Since the good performance of our model depends on the partial output of the previous moment, although we introduce some blank frames for the first few frames, the lack of information causes the metrics of the first few frames to be not very satisfactory.So our future work will focus on this part.
\bibliographystyle{IEEEbib}
\bibliography{icme2023template}
\end{document}